%% file: geometric.tex
\documentclass[onecolumn,12pt,conference]{IEEEtran}
\pdfoutput=1
\usepackage[letterpaper,text={6in,9in},left=1.25in,top=1in]{geometry}
\usepackage[font=small,format=hang]{caption}
\usepackage{amsmath}
\usepackage{amsthm}
\usepackage{amssymb}
\usepackage{bm}
\usepackage{xspace}
\usepackage{xcolor}
\usepackage{graphicx}
\usepackage{url}
\usepackage{cite}
\usepackage{algorithm}
\usepackage{algorithmicx}
\usepackage{algpseudocode}
\usepackage{float}

\theoremstyle{plain}
\newtheorem{proposition}{Proposition}
\newtheorem{lemma}{Lemma}
\theoremstyle{definition}

\theoremstyle{remark}

\newcommand{\vecp}{\boldsymbol{p}}
\newcommand{\vecq}{\boldsymbol{q}}
\newcommand{\vecr}{\boldsymbol{r}}
\newcommand{\vecu}{\boldsymbol{u}}
\newcommand{\vecw}{\boldsymbol{w}}
\newcommand{\vecx}{\boldsymbol{x}}

\DeclareMathOperator*{\maximize}{maximize}

\DeclareMathOperator*{\argmax}{argmax}

\DeclareMathOperator{\kl}{D}
\DeclareMathOperator{\entop}{\mathcal{H}}
\DeclareMathOperator{\miop}{\mathcal{I}}

\IEEEoverridecommandlockouts

\title{Matching Dyadic Distributions to Channels}

\author{\begin{minipage}{\textwidth}\normalsize\begin{center}G. B\"ocherer and R. Mathar\\
Institute for Theoretical Information
Technology\\
RWTH Aachen University, 52056 Aachen, Germany\\
Email: \texttt{\{boecherer,mathar\}@ti.rwth-aachen.de}
\thanks{This work has been supported by the UMIC Research Centre, RWTH
Aachen University.
}
\end{center}
\end{minipage}
}

\def\thebibliography#1{\section*{References}\normalsize\list
    {[\arabic{enumi}]}{\settowidth\labelwidth{[#1]}\leftmargin\labelwidth
    \advance\leftmargin\labelsep \itemsep 0pt plus .5pt
    \usecounter{enumi}}
    \def\newblock{\hskip .11em plus .33em minus .07em}
    \sloppy\clubpenalty4000\widowpenalty4000
    \sfcode`\.=1000\relax}

\begin{document}
\maketitle

\input{abstract}

\input{introduction}

\input{algo}

\input{dmc}

\input{dnc}

\appendix

\input{ghc}

\bibliographystyle{IEEEtran}
\normalsize
\bibliography{IEEEabrv,confs-jrnls,Dnc}

\end{document}

%% file: abstract.tex
\begin{abstract}
\normalsize Many communication channels with discrete input have non-uniform capacity achieving probability mass functions (PMF). By parsing a stream of independent and equiprobable bits according to a full prefix-free code, a modu-lator can generate dyadic PMFs at the channel input. In this work, we show that for discrete memoryless channels and for memoryless discrete noiseless channels, searching for good dyadic input PMFs is equivalent to minimizing the Kullback-Leibler distance between a dyadic PMF and a weighted version of the capacity achieving PMF. We define a new algorithm called Geometric Huffman Coding (GHC) and prove that GHC finds the optimal dyadic PMF in $\mathcal{O}(m \log m)$ steps where $m$ is the number of input symbols of the considered channel. Furthermore, we prove that by generating dyadic PMFs of blocks of consecutive input symbols, GHC achieves capacity when the block length goes to infinity.
\end{abstract}

%% file: introduction.tex
\section{introduction}

For many communication channels, the ultimate rate for reliable data transmission is given by the maximum information per cost. For discrete memoryless channels (DMC) and for additive noise channels with finite input alphabet, the ultimate rate is the maximum mutual information between input and output per channel use. For memoryless discrete noiseless channels (DNC), the ultimate rate is the maximum entropy of the input per average weight. In both cases, the maximum is achieved by an input that is distributed according to a capacity achieving probability mass function (PMF). To use non-uniform input PMFs in a digital communication system, a modulator has to generate this PMF by mapping independent equiprobable data bits to the channel input symbols. One way to do this is to parse the data bits by a full prefix-free code and to map each codeword to an input symbol \cite[Sec.~VII]{Kschischang1993}. PMFs that can be generated in this way are dyadic, i.e., the probability of each point is of the form $2^{-\ell},\ell\in\mathbb{N}$. The capacity achieving PMFs are in general not dyadic, which raises two questions. First, what is an optimal dyadic PMF that maximizes information per cost, and second, if we jointly generate blocks of consecutive input symbols by a dyadic PMF, can we asymptotically achieve capacity by letting the block length go to infinity.

For noiseless channels, an efficient algorithm to find the optimal dyadic PMF that maximizes entropy per average weight was found in \cite{Lempel1973}. In general, a common approach in the literature is to use the dyadic PMF that results from the optimal \emph{source code} of the capacity achieving PMF. Dyadic PMFs resulting from source codes are in general not optimal. For the $(d,k)$ constrained noiseless channel, it was claimed in \cite{Kerpez1991} that a source code \emph{asymptotically} achieves capacity. To the best of our knowledge, for DMCs, there exist no results in the literature on optimality and asymptotic behavior of dyadic PMFs. In \cite{Kschischang1993,Ungerboeck2002}, the authors use source codes for additive noise channels. While good numerical results are observed, optimality and asymptotic behavior are not assessed. In \cite{Abrahams1998}, input entropy per average weight is maximized for additive noise channels. This is in general not equivalent to the maximization of mutual information per channel use.

Denote the capacity achieving PMF of a channel by $\vecp^*$. In this work, we show for DMCs that minimizing the Kullback-Leibler distance (KL) $\kl(\vecp\Vert\vecp^*)$ over all dyadic PMFs $\vecp$ maximizes a lower bound on the achieved mutual information per channel use. For DNCs, we show that searching for the optimal dyadic input PMF is equivalent to minimizing the weighted  KL-distance \text{$\kl(\vecp\Vert\vecp^{*\mathsf{R}})\triangleq\sum_i p_i \log(p_i/{p_i^*}^\mathsf{R})$} over all dyadic PMFs $\vecp$. The value of $\mathsf{R}$ is given by the fraction of the channel capacity that is achievable by dyadic PMFs. We introduce an algorithm called Geometric Huffman Coding (\textsc{Ghc}) and prove that \textsc{Ghc} minimizes $\kl(\vecp\Vert\vecx)$ over all dyadic PMFs $\vecp$, for any given vector $\vecx$ with non-negative entries. In particular, for $\vecx=\vecp^*$, \textsc{Ghc} minimizes $\kl(\vecp\Vert\vecp^*)$ and for $\vecx=\vecp^{*\mathsf{R}}$, \textsc{Ghc} minimizes $\kl(\vecp\Vert\vecp^{*\mathsf{R}})$. The complexity of \textsc{Ghc} is $\mathcal{O}(m\log m)$, where $m$ is the number of input symbols of the considered channel. Furthermore, we show that, to asymptotically achieve capacity for DMCs and DNCs, the normalized KL-distance $\kl(\vecp^{(k)}\Vert\vecp^{(k)*})/k$ has to vanish for block length $k\rightarrow\infty$. This is achieved by \textsc{Ghc}. Based on the present work, we show in \cite{Bocherer2010e} that for finite signal constellations with average power constraint, \textsc{Ghc} achieves capacity. \textsc{Ghc} is as handy as Huffman coding and an implementation of \textsc{Ghc} in MATLAB is readily available at our website \cite{website:ghc}.

The remainder of this work is organized as follows. In Section~\ref{sec:ghc}, we define \textsc{Ghc}. In Section~\ref{sec:dmc}, we show optimality and asymptotic optimality of \textsc{Ghc} for DMCs. We show optimality and asymptotic optimality of \textsc{Ghc} for DNCs in Section~\ref{sec:dnc}.

%% file: algo.tex
\section{Geometric Huffman Coding}
\label{sec:ghc}
For a PMF $\vecp$ and a vector $\vecx$ with non-negative entries, the KL-distance is given by
\begin{align}
\kl(\vecp\Vert\vecx)=\sum_i p_i\log \frac{p_i}{x_i}.\label{eq:wKL}
\end{align}
Note that $\kl(\vecp\Vert\vecx)$ can be equal to infinity. The dyadic PMF $\vecp$ that minimizes the KL-distance is directly given by the full prefix-free code that is constructed by the algorithm of the following proposition. A prefix-free code is full if it fulfills the Kraft inequality \cite[Theorem~5.2.2]{Cover2006} with equality.
\begin{proposition}\label{prop:ghc}
Without loss of generality, we assume $x_1\geq x_2\geq \dotsb \geq x_m$. The dyadic PMF $\vecp$ that minimizes $\kl(\vecp\Vert\vecx)$ is obtained by constructing a Huffman tree with the updating rule
\begin{align}
x'=
\left\{
\begin{array}{ll}
x_{m-1},&\text{if } x_{m-1}\geq 4 x_m\\
2\sqrt{x_{m-1}x_m},&\text{if }x_{m-1}< 4 x_m.
\end{array}
\right.
\end{align}
Since it involves a geometric mean, we call this method \emph{Geometric Huffman Coding}. We write $\vecp=\textsc{Ghc}(\vecx)$.
\end{proposition}
\begin{IEEEproof}
The proof is given in the appendix.
\end{IEEEproof}
An implementation of \textsc{Ghc} in MATLAB can be found at our website \cite{website:ghc}. In comparison to \textsc{Ghc}, Huffman coding uses the updating rule $x'=x_m+x_{m-1}$. Furthermore, it can be shown that Huffman coding minimizes the KL-distance $\kl(\vecx\Vert\vecp)$ over all dyadic PMFs $\vecp$. Note that this is not equivalent to minimizing \eqref{eq:wKL} because the KL-distance is not symmetric in its arguments. \textsc{Ghc} has the same complexity as Huffman coding, which is $\mathcal{O}(m\log m)$ \cite[Chap. 16.3]{Cormen2001}.
\begin{figure}
\centering
\def\svgwidth{1.0\textwidth}
\footnotesize
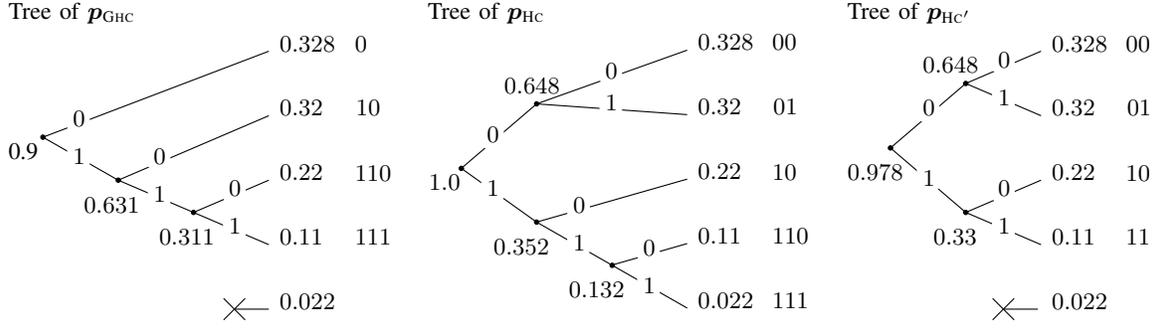
\caption{For $\vecq=(0.328,0.32,0.22,0.11,0.022)^T$, the left figure displays the code tree of \textsc{Ghc}. The figure in the middle shows the code tree of Huffman coding. The right figure displays the code tree of Huffman coding applied to $(q_1,\dotsc,q_4)^T$.}
\label{fig:example}
\end{figure}

For illustration purpose, we apply \textsc{Ghc} and Huffman coding to the PMF
\begin{align}
\vecq=(0.328,0.32,0.22,0.11,0.022)^T
\end{align}
where $(\cdot)^T$ denotes the transpose. The resulting code trees are displayed in Figure~\ref{fig:example}. By reading off the codeword lengths, the corresponding dyadic PMFs are
\begin{align}
\vecp_\textsc{Ghc}=(2^{-1},2^{-2},2^{-3},2^{-3},0)^T\text{ and }\vecp_\textsc{Hc}=(2^{-2},2^{-2},2^{-2},2^{-3},2^{-3})^T
\end{align}
and the KL-distances to $\vecq$ are
\begin{align}
\kl(\vecp_\textsc{Ghc}\Vert\vecq)=0.13619\text{ and }\kl(\vecp_\textsc{Hc}\Vert\vecq)=0.19548
\end{align}
where we used the dual logarithm. As expected, the KL-distance resulting from \textsc{Ghc} is smaller than the one that results from Huffman coding. Since \textsc{Ghc} assigns zero to $q_5$, one may want to manually assign probability zero to $q_5$ and then apply Huffman coding to $(q_1,\dotsc,q_4)^T$. The corresponding code tree is displayed in Figure~\ref{fig:example}. The corresponding PMF and the resulting KL-distance to $\vecq$ are respectively
\begin{align}
\vecp_{\textsc{Hc}'}=(2^{-2},2^{-2},2^{-2},2^{-2},0)^T,\quad\kl(\vecp_{\textsc{Hc}'}\Vert\vecq)=0.15523.
\end{align}
While $\vecp_{\textsc{Hc}'}$ slightly improves upon $\vecp_{\textsc{Hc}}$, the KL-distance is still larger than the one resulting from \textsc{Ghc}.

Let $\vecq$ now denote some arbitrary PMF. We consider $k$ subsequent symbols that are independent and identically distributed according to $\vecq$. We denote the joint PMF of these symbols by $\vecq^{(k)}$. Our aim is to show that for $\vecp^{(k)}=\textsc{Ghc}(\vecq^{(k)})$, the normalized KL-distance $\kl(\vecp^{(k)}\Vert\vecq^{(k)})/k$ vanishes for $k\rightarrow\infty$. To show this, we will need the following lemma, which shows the existence of dyadic PMFs with a bounded KL-distance for any PMF $\vecq$.
\begin{lemma}\label{lem:gcc}
Without loss of generality, $q_1\geq q_2\geq\dotsb \geq q_m$. Assign then $p_i=2^{-\lfloor -\log_2 q_i\rfloor}$ for $i\leq k$, and $p_i=0$ for $i>k$, where $k\leq m$ is chosen\footnote{It can actually be shown that such $k$ always exists, so \textsc{Gcc} is well-defined.} such that $\sum_{i=1}^m p_i = 1$. Then $\vecp$ is a dyadic PMF and $\kl(\vecp\Vert\vecq)\leq \log 2$. We call this method \emph{Greedy Channel Coding} (\textsc{Gcc}) and write $\vecp=\textsc{Gcc}(\vecq)$.
\end{lemma}
\begin{IEEEproof}
\begin{align}
\kl(\vecp\Vert\vecq)=\sum\limits_{i=1}^{m} p_i\log\frac{p_i}{q_i}\label{eq:gccOne}
&\leq\sum\limits_{i=1}^m p_i\log\frac{2^{-\lfloor -\log_2 q_i\rfloor}}{q_i}\\
&\leq\sum\limits_{i=1}^m p_i\log\frac{2^{-( -\log_2 q_i-1)}}{q_i}\\
&=\sum\limits_{i=1}^m p_i \log\frac{2q_i}{q_i}=\log 2
\end{align}
where the inequality in \eqref{eq:gccOne} follows from the values that \textsc{Gcc} assigns to the $p_i$.
\end{IEEEproof}
An implementation of \textsc{Gcc} in MATLAB is available at \cite{website:ghc}. It is now easy to show the asymptotic behavior of \textsc{Ghc}.
\begin{proposition}
\label{prop:asymptotic}
For $\vecp^{(k)}=\textsc{Ghc}(\vecq^{(k)})$ it holds that
\begin{align}
\frac{\kl(\vecp^{(k)}\Vert \vecq^{(k)})}{k}\overset{k\rightarrow\infty}{\longrightarrow}0.
\end{align}
\end{proposition}
\begin{IEEEproof}
Define $\tilde{\vecp}^{(k)}=\textsc{Gcc}(\vecq^{(k)})$. Then
\begin{align}
\frac{\kl(\vecp^{(k)}\Vert \vecq^{(k)})}{k}&\leq \frac{\kl(\tilde{\vecp}^{(k)}\Vert \vecq^{(k)})}{k}\leq\frac{\log 2}{k}
\end{align}
where the first inequality follows via Proposition~\ref{prop:ghc} from the optimality of \textsc{Ghc} and where the second inequality follows from Lemma~\ref{lem:gcc}. $\log 2/k$ goes to zero for $k\rightarrow\infty$ and the statement of the proposition follows.
\end{IEEEproof}

%% file: example.pdf_tex

\begingroup
  \makeatletter
  \providecommand\color[2][]{%
    \errmessage{(Inkscape) Color is used for the text in Inkscape, but the package 'color.sty' is not loaded}
    \renewcommand\color[2][]{}%
  }
  \providecommand\transparent[1]{%
    \errmessage{(Inkscape) Transparency is used (non-zero) for the text in Inkscape, but the package 'transparent.sty' is not loaded}
    \renewcommand\transparent[1]{}%
  }
  \providecommand\rotatebox[2]{#2}
  \ifx\svgwidth\undefined
    \setlength{\unitlength}{852.6953125pt}
  \else
    \setlength{\unitlength}{\svgwidth}
  \fi
  \global\let\svgwidth\undefined
  \makeatother
  \begin{picture}(1,0.27183208)%
    \put(0,0){\includegraphics[width=\unitlength]{example.pdf}}%
    \put(0.2361629,0.2348849){\color[rgb]{0,0,0}\makebox(0,0)[lb]{\smash{$0.328$}}}%
    \put(0.2361629,0.17859281){\color[rgb]{0,0,0}\makebox(0,0)[lb]{\smash{$0.32$}}}%
    \put(0.2361629,0.12230072){\color[rgb]{0,0,0}\makebox(0,0)[lb]{\smash{$0.22$}}}%
    \put(0.2361629,0.06600864){\color[rgb]{0,0,0}\makebox(0,0)[lb]{\smash{$0.11$}}}%
    \put(0.2361629,0.00971655){\color[rgb]{0,0,0}\makebox(0,0)[lb]{\smash{$0.022$}}}%
    \put(0.60112328,0.2358231){\color[rgb]{0,0,0}\makebox(0,0)[lb]{\smash{$0.328$}}}%
    \put(0.60112328,0.17953101){\color[rgb]{0,0,0}\makebox(0,0)[lb]{\smash{$0.32$}}}%
    \put(0.60112328,0.12323893){\color[rgb]{0,0,0}\makebox(0,0)[lb]{\smash{$0.22$}}}%
    \put(0.60112328,0.06694684){\color[rgb]{0,0,0}\makebox(0,0)[lb]{\smash{$0.11$}}}%
    \put(0.60112328,0.01065475){\color[rgb]{0,0,0}\makebox(0,0)[lb]{\smash{$0.022$}}}%
    \put(0.90979156,0.2348849){\color[rgb]{0,0,0}\makebox(0,0)[lb]{\smash{$0.328$}}}%
    \put(0.90979156,0.17859281){\color[rgb]{0,0,0}\makebox(0,0)[lb]{\smash{$0.32$}}}%
    \put(0.90979156,0.12230072){\color[rgb]{0,0,0}\makebox(0,0)[lb]{\smash{$0.22$}}}%
    \put(0.90979156,0.06600864){\color[rgb]{0,0,0}\makebox(0,0)[lb]{\smash{$0.11$}}}%
    \put(0.90979156,0.00971655){\color[rgb]{0,0,0}\makebox(0,0)[lb]{\smash{$0.022$}}}%
    \put(0.30183701,0.2348849){\color[rgb]{0,0,0}\makebox(0,0)[lb]{\smash{$0$}}}%
    \put(0.30277521,0.17859281){\color[rgb]{0,0,0}\makebox(0,0)[lb]{\smash{$10$}}}%
    \put(0.30183701,0.12136252){\color[rgb]{0,0,0}\makebox(0,0)[lb]{\smash{$110$}}}%
    \put(0.30183701,0.06600864){\color[rgb]{0,0,0}\makebox(0,0)[lb]{\smash{$111$}}}%
    \put(0.66585918,0.2358231){\color[rgb]{0,0,0}\makebox(0,0)[lb]{\smash{$00$}}}%
    \put(0.66679738,0.17953101){\color[rgb]{0,0,0}\makebox(0,0)[lb]{\smash{$01$}}}%
    \put(0.66585918,0.12230072){\color[rgb]{0,0,0}\makebox(0,0)[lb]{\smash{$10$}}}%
    \put(0.66585918,0.06694684){\color[rgb]{0,0,0}\makebox(0,0)[lb]{\smash{$110$}}}%
    \put(0.66585918,0.01065475){\color[rgb]{0,0,0}\makebox(0,0)[lb]{\smash{$111$}}}%
    \put(0.97452746,0.2348849){\color[rgb]{0,0,0}\makebox(0,0)[lb]{\smash{$00$}}}%
    \put(0.97546566,0.17859281){\color[rgb]{0,0,0}\makebox(0,0)[lb]{\smash{$01$}}}%
    \put(0.97452746,0.12136252){\color[rgb]{0,0,0}\makebox(0,0)[lb]{\smash{$10$}}}%
    \put(0.97452746,0.06600864){\color[rgb]{0,0,0}\makebox(0,0)[lb]{\smash{$11$}}}%
    \put(0.13108434,0.06600864){\color[rgb]{0,0,0}\makebox(0,0)[lb]{\smash{$0.311$}}}%
    \put(0.06541023,0.09415468){\color[rgb]{0,0,0}\makebox(0,0)[lb]{\smash{$0.631$}}}%
    \put(-0.00026387,0.14106475){\color[rgb]{0,0,0}\makebox(0,0)[lb]{\smash{0.9}}}%
    \put(0.4885391,0.02003676){\color[rgb]{0,0,0}\makebox(0,0)[lb]{\smash{$0.132$}}}%
    \put(0.422865,0.05756482){\color[rgb]{0,0,0}\makebox(0,0)[lb]{\smash{$0.352$}}}%
    \put(0.36657291,0.11385691){\color[rgb]{0,0,0}\makebox(0,0)[lb]{\smash{$1.0$}}}%
    \put(0.43224701,0.19829504){\color[rgb]{0,0,0}\makebox(0,0)[lb]{\smash{$0.648$}}}%
    \put(0.8065894,0.06600864){\color[rgb]{0,0,0}\makebox(0,0)[lb]{\smash{$0.33$}}}%
    \put(0.79720738,0.21612087){\color[rgb]{0,0,0}\makebox(0,0)[lb]{\smash{$0.648$}}}%
    \put(0.73153328,0.12230072){\color[rgb]{0,0,0}\makebox(0,0)[lb]{\smash{$0.978$}}}%
    \put(0.05602822,0.1692108){\color[rgb]{0,0,0}\makebox(0,0)[lb]{\smash{$0$}}}%
    \put(0.05602822,0.13637375){\color[rgb]{0,0,0}\makebox(0,0)[lb]{\smash{$1$}}}%
    \put(0.12639333,0.13637375){\color[rgb]{0,0,0}\makebox(0,0)[lb]{\smash{$0$}}}%
    \put(0.19206743,0.1082277){\color[rgb]{0,0,0}\makebox(0,0)[lb]{\smash{$0$}}}%
    \put(0.12639333,0.10353669){\color[rgb]{0,0,0}\makebox(0,0)[lb]{\smash{$1$}}}%
    \put(0.19206743,0.07539065){\color[rgb]{0,0,0}\makebox(0,0)[lb]{\smash{$1$}}}%
    \put(0.41723579,0.15513778){\color[rgb]{0,0,0}\makebox(0,0)[lb]{\smash{$0$}}}%
    \put(0.52043795,0.21142986){\color[rgb]{0,0,0}\makebox(0,0)[lb]{\smash{$0$}}}%
    \put(0.49229191,0.09415468){\color[rgb]{0,0,0}\makebox(0,0)[lb]{\smash{$0$}}}%
    \put(0.553275,0.05662662){\color[rgb]{0,0,0}\makebox(0,0)[lb]{\smash{$0$}}}%
    \put(0.79720738,0.17859281){\color[rgb]{0,0,0}\makebox(0,0)[lb]{\smash{$0$}}}%
    \put(0.86288149,0.22081188){\color[rgb]{0,0,0}\makebox(0,0)[lb]{\smash{$0$}}}%
    \put(0.86288149,0.1082277){\color[rgb]{0,0,0}\makebox(0,0)[lb]{\smash{$0$}}}%
    \put(0.52043795,0.18328382){\color[rgb]{0,0,0}\makebox(0,0)[lb]{\smash{$1$}}}%
    \put(0.41723579,0.1082277){\color[rgb]{0,0,0}\makebox(0,0)[lb]{\smash{$1$}}}%
    \put(0.49229191,0.06131763){\color[rgb]{0,0,0}\makebox(0,0)[lb]{\smash{$1$}}}%
    \put(0.553275,0.02378957){\color[rgb]{0,0,0}\makebox(0,0)[lb]{\smash{$1$}}}%
    \put(0.86288149,0.18797483){\color[rgb]{0,0,0}\makebox(0,0)[lb]{\smash{$1$}}}%
    \put(0.79720738,0.11760972){\color[rgb]{0,0,0}\makebox(0,0)[lb]{\smash{$1$}}}%
    \put(0.86288149,0.07539065){\color[rgb]{0,0,0}\makebox(0,0)[lb]{\smash{$1$}}}%
    \put(-0.00026387,0.26303095){\color[rgb]{0,0,0}\makebox(0,0)[lb]{\smash{Tree of $\vecp_\textsc{Ghc}$}}}%
    \put(0.36563471,0.26303095){\color[rgb]{0,0,0}\makebox(0,0)[lb]{\smash{Tree of $\vecp_\textsc{Hc}$}}}%
    \put(0.73153328,0.26303095){\color[rgb]{0,0,0}\makebox(0,0)[lb]{\smash{Tree of $\vecp_{\textsc{Hc}'}$}}}%
  \end{picture}%
\endgroup

%% file: dmc.tex
\section{Discrete Memoryless Channel}
\label{sec:dmc}
We now show how \textsc{Ghc} can be used to find dyadic PMFs that well approximate the capacity of DMCs. A DMC is specified by a set of $m$ input symbols, a set of $n$ output symbols and a matrix of transition probabilities $(h_{ji})$. An input PMF $\vecp$ relates to its corresponding output PMF $\vecr$ as
\begin{align}
\vecr=
\left(
\begin{array}{c}
r_1\\
\vdots\\
r_n
\end{array}
\right)
=
\left(
\begin{array}{ccc}
h_{11}&\cdots&h_{1m}\\
\vdots&\ddots&\vdots\\
h_{n1}&\cdots&h_{nm}
\end{array}
\right)
\left(
\begin{array}{c}
p_1\\
\vdots\\
p_m
\end{array}
\right)
.
\end{align}
The mutual information between input and output is given by \cite[Eq. (8.73)]{Gallager2008}
\begin{align}
\miop(\vecp)&=\sum_i p_i\sum_j h_{ji}\log\frac{h_{ji}}{r_j}.
\end{align}
The capacity of a DMC is the maximum mutual information between input and output, where the maximum is taken over all input PMFs. To find the best dyadic input PMF, we need to solve the optimization problem
\begin{align}
\maximize_{\vecp} \quad&\miop(\vecp)\nonumber\\
\mathrm{subject\,to} \quad& \vecp\text{ is a PMF}\nonumber\\
& p_i = 2^{-\ell_i},\ell_i\in\mathbb{N},i=1,\dotsc,m.
\end{align}
This is a nonlinear optimization problem with integer constraints and therefore intractable for practical purposes. In order to overcome this difficulty, we proceed as follows. First, we will drop the restriction to dyadic PMFs and characterize the capacity achieving PMF $\vecp^*$. Then, we will derive the penalty that results from using a PMF $\vecp$ different from $\vecp^*$. Finally, we will minimize this penalty over all dyadic PMFs.

Capacity and capacity achieving PMF are respectively defined as
\begin{align}
\mathsf{C}=\max_{\vecp}\miop(\vecp),\qquad\vecp^*=\argmax_{\vecp} \miop(\vecp).
\end{align}
Denote by $\vecr$ and $\vecr^*$ the output PMFs that result from using the input PMFs $\vecp$ and $\vecp^*$, respectively. According to \cite[Eq.~(4.5.1)]{Gallager1968}, the output PMF $\vecr^*$ resulting from the capacity achieving PMF $\vecp^*$ has the important property that
\begin{align}
\sum_j h_{ji}\log\frac{h_{ji}}{r^*_j}=\mathsf{C},\quad\text{whenever $p^*_i>0$}.\label{eq:dmcKKT}
\end{align}
We now use this property to express the mutual information $\miop(\vecp)$ achieved by some PMF $\vecp$ in terms of capacity $\mathsf{C}$ and capacity achieving PMF $\vecp^*$. The only assumption that we make about $\vecp$ is that
\begin{align}
p_i=0,\quad\text{whenever $p_i^*=0$}.\label{eq:dmcCondition}
\end{align}
Under this assumption, we have for $\miop(\vecp)$
\begin{align}
\miop(\vecp)=\sum_i p_i\sum_j h_{ji}\log\frac{h_{ji}}{r_j}
&=\sum_i p_i\sum_j h_{ji}\log\frac{h_{ji}r_j^*}{r_jr_j^*}\\
&=\sum_i p_i\sum_j h_{ji}\log\frac{h_{ji}}{r_j^*}+\sum_i p_i\sum_j h_{ji}\log\frac{r_j^*}{r_j}\\
&=\mathsf{C}-\sum_j \Bigl(\sum_i p_ih_{ji}\Bigr)\log\frac{r_j}{r_j^*}\label{eq:dmcTrick}\\
&=\mathsf{C}-\sum_j r_j\log\frac{r_j}{r_j^*}\\
&=\mathsf{C}-\kl(\vecr\Vert\vecr^*)
\end{align}
where equality in \eqref{eq:dmcTrick} follows from \eqref{eq:dmcKKT} and \eqref{eq:dmcCondition}. From the last line, we see that the penalty of using $\vecp$ instead of $\vecp^*$ is exactly the KL-distance between the corresponding output PMFs $\vecr$ and $\vecr^*$. To get a simple expression that directly depends on $\vecp$ and $\vecp^*$, we lower bound the last line. According to \cite[Eq.~(4.45)]{Cover2006} the KL-distance between the output PMFs is upper-bounded by the KL-distance between the input PMFs, i.e, $\kl(\vecr\Vert\vecr^*)\leq \kl(\vecp\Vert\vecp^*)$.
Thus, 
\begin{align}
\miop(\vecp)\geq \mathsf{C}-\kl(\vecp\Vert\vecp^{*}).\label{eq:dmcBound}
\end{align}
We conclude that for DMCs, the penalty that results from using $\vecp$ instead of $\vecp^*$ is upper bounded by $\kl(\vecp\Vert\vecp^*)$. According to Proposition~\ref{prop:ghc}, we can now efficiently minimize the penalty bound over all dyadic input PMFs $\vecp$ by using $\vecp=\textsc{Ghc}(\vecp^*)$. Note that \textsc{Ghc} guarantees \eqref{eq:dmcCondition}: assume $\vecp^*$ is ordered and $p^*_m=0$, $p^*_{m-1}>0$. Then $p^*_{m-1}>4p^*_m$ and \textsc{Ghc} assigns $p_m=0$.

We now jointly consider the PMF of $k$ consecutive channel inputs. Denote by $\vecp^{(k)*}$ the capacity achieving joint PMF. Since the channel is memoryless, $\vecp^{(k)*}$ is the product of $k$ marginal PMFs $\vecp^*$ and we have $\miop(\vecp^{(k)*})=k\miop(\vecp^*)=k\mathsf{C}$. Thus, for a joint PMF $\vecp^{(k)}$ we have
\begin{align}
\miop(\vecp^{(k)})\geq k\mathsf{C}-\kl(\vecp^{(k)}\Vert\vecp^{(k)*}).
\end{align} 
The mutual information per channel use $\bar{\miop}(\vecp^{(k)})\triangleq\miop(\vecp^{(k)})/k$ is thus given by
\begin{align}
\bar{\miop}(\vecp^{(k)})\geq \mathsf{C}-\frac{\kl(\vecp^{(k)}\Vert\vecp^{(k)*})}{k}.
\end{align}
By using $\vecp^{(k)}=\textsc{Ghc}(\vecp^{(k)*})$, according to Proposition~\ref{prop:asymptotic}, $\bar{\miop}(\vecp^{(k)})\to\mathsf{C}$ for $k\rightarrow\infty$ and we conclude that \textsc{Ghc} is asymptotically capacity achieving.

%% file: dnc.tex
\section{Memoryless Discrete Noiseless Channel}
\label{sec:dnc}
Following \cite{Krause1962}, a memoryless DNC is given by a finite alphabet $\mathcal{A}=(a_1,\dotsc,a_m)$ of atomic symbols and an associated weight function $w\colon\mathcal{A}\rightarrow\mathbb{R}_{>0}$, $a_i\mapsto w_i>0$.  
The information rate $\bar{\entop}$ that is transmitted over the channel is given by the entropy of the input PMF divided by the average weight, i.e.,
\begin{align}
\bar{\entop}(\vecp)=\frac{\entop(\vecp)}{\sum_i p_iw_i},\qquad \text{with } \entop(\vecp)=-\sum_i p_i \log p_i.
\end{align}
To find the dyadic PMF that maximizes $\bar{\entop}(\vecp)$, we need to solve the optimization problem
\begin{align}
\maximize_{\vecp}\quad&\bar{\entop}(\vecp)\nonumber\\
\mathrm{subject\,to}\quad&\text{$\vecp$ is a PMF}\nonumber\\
&p_i=2^{-\ell_i},\ell_i\in\mathbb{N},i=1,\dotsc,m.
\end{align}
As in the case of DMCs, this is an intractable nonlinear optimization problem with integer constraints. We will therefore proceed in the same way as we did for the DMC in Section~\ref{sec:dmc}.
We start by calculating the capacity and the capacity achieving PMF $\vecp^*$. This can be done by Lagrange Multipliers, see, e.g., \cite{Marcus1957}. Denote by $b$ the base of the logarithm $\log$. The capacity is achieved by the input PMF
\begin{align}
p^*_i=b^{-\mathsf{C}w_i},\quad i=1,\dotsc,m\label{eq:dncCAD}
\end{align}
where $\mathsf{C}$ denotes capacity and is given by the greatest positive real solution of the equation
\begin{align}
\sum\limits_i b^{-sw_i}=1.
\end{align}
From \eqref{eq:dncCAD}, we have the relation $w_i=-\frac{1}{\mathsf{C}}\log p^*_i$. We can thus write
\begin{align}
\sum\limits_i p_i w_i=-\frac{1}{\mathsf{C}}\sum\limits_i p_i \log p_i^*.\label{eq:weightPMF}
\end{align}
Denote by $\mathsf{R}$ the fraction of $\mathsf{C}$ that can be achieved by the best dyadic PMF $\tilde{\vecp}$, i.e.,
\begin{align}
\tilde{\vecp}\triangleq\argmax_{\vecp\text{ dyadic}}\bar{\entop}(\vecp),\qquad \mathsf{R} \triangleq\frac{\bar{\entop}(\tilde{\vecp})}{\mathsf{C}}.\label{eq:dncDyadic}
\end{align}
In general, $\mathsf{R}$ is not known beforehand, but we will show in Subsection~\ref{subsec:lec} how it can be found. Suppose for now that we know $\mathsf{R}$. Assume further that
\begin{align}
p_i=0,\quad\text{whenever } p_i^* = 0.\label{eq:dncCondition}
\end{align}
Furthermore, we use the convention $0\log 0 =0$. With these assumptions, we can now write $\bar{\entop}(\vecp)$ as
\begin{align}
\bar{\entop}(\vecp)&=\frac{-\mathsf{R}\sum_i p_i\log p_i^*+\mathsf{R}\sum_i p_i\log p_i^*+\entop(\vecp)}{\sum\limits_i p_i w_i}\\
&=\mathsf{R}\mathsf{C}-\frac{\sum_i p_i\log p_i-\mathsf{R}\sum_i p_i\log p_i^*}{\sum\limits_i p_i w_i}\label{eq:dncPenaltyTwo}\\
&=\mathsf{R}\mathsf{C}-\frac{\sum_i p_i\log \frac{p_i}{{p_i^*}^\mathsf{R}}}{\sum\limits_i p_i w_i}
=\mathsf{R}\mathsf{C}-\frac{\kl(\vecp\Vert\vecp^{*\mathsf{R}})}{\sum_ip_iw_i}\label{eq:dncPenalty}
\end{align}
where in \eqref{eq:dncPenaltyTwo}, we used \eqref{eq:weightPMF} and the definition of entropy. By \eqref{eq:dncDyadic}, for the best dyadic PMF $\tilde{\vecp}$ we have $\bar{\entop}(\tilde{\vecp})=\mathsf{R}\mathsf{C}$. It follows that for any dyadic PMF $\vecp$, we have $\kl(\vecp\Vert\vecp^{*\mathsf{R}})\geq 0$ and for the best dyadic PMF $\tilde{\vecp}$, we have $\kl(\tilde{\vecp}\Vert\vecp^{*\mathsf{R}})=0$. We conclude that for DNCs, the best dyadic PMF is found by minimizing $\kl(\vecp\Vert\vecp^{*\mathsf{R}})$ over all dyadic PMFs $\vecp$ and by Proposition~\ref{prop:ghc}, this PMF is given by $\vecp=\textsc{Ghc}(\vecp^{*\mathsf{R}})$. Recall that, as we argued in Section~\ref{sec:dmc}, \textsc{Ghc} guarantees \eqref{eq:dncCondition}.

We now consider the PMF of $k$ consecutive symbols. We denote the corresponding weights by $\vecw^{(k)}$. The capacity achieving joint PMF is the product of $k$ copies of $\vecp^*$ and we denote it by $\vecp^{(k)*}$. Clearly, $w^{(k)}_i\geq kw_{\min}$ for $i=1,\dotsc,m^k$ where $w_{\min}=\min\{w_1,\dotsc,w_m\}$. Using this, we get for $\bar{\entop}(\vecp^{(k)})$ the lower bound
\begin{align}
\bar{\entop}(\vecp^{(k)})=\frac{\entop(\vecp^{(k)})}{\sum_ip^{(k)}_iw_i^{(k)}}
&=\frac{\entop(\vecp^{(k)})+\sum_ip^{(k)}_i\log p_i^{(k)*}-\sum_ip^{(k)}_i\log p_i^{(k)*}}{-\frac{1}{\mathsf{C}}\sum_ip^{(k)}_i\log p_i^{(k)*}}\\
&=\mathsf{C}-\frac{\kl(\vecp^{(k)}\Vert\vecp^{(k)*})}{\sum_ip^{(k)}_iw_i^{(k)}}\\
&\geq \mathsf{C}-w_{\min}^{-1}\frac{\kl(\vecp^{(k)}\Vert\vecp^{(k)*})}{k}.
\end{align}
For $\vecp^{(k)}=\textsc{Ghc}(\vecp^{(k)*})$, according to Proposition~\ref{prop:asymptotic}, the last term in the last line vanishes for $k\rightarrow\infty$ and we have $\bar{\entop}(\vecp^{(k)})\rightarrow\mathsf{C}$, thus \textsc{Ghc} is asymptotically capacity achieving.

\subsection{Finding $\mathsf{R}$}
\label{subsec:lec}
\begin{algorithm}[t]
\caption{Finding $\mathsf{R}$ and the optimal dyadic PMF for DNCs}
\label{alg:lec}
\begin{algorithmic}[1]
\Procedure{Lec}{$\vecp^*$}
\State $R\leftarrow 1$
\While{$\kl(\vecp\Vert\vecp^{*R})\neq 0$}
\State $\vecp\leftarrow \textsc{Ghc}(\vecp^{*R})$
\State $R\leftarrow\bar{\entop}(\vecp)/\mathsf{C}$
\EndWhile
\EndProcedure
\end{algorithmic}
\end{algorithm}

The exact value of $\mathsf{R}$ is in general not known beforehand. However, $\mathsf{R}$ and the best dyadic PMF can be found iteratively by the Lempel-Even-Cohn (\textsc{Lec}) algorithm \cite{Lempel1973}. The idea of the algorithm is to start with some $R$, then find the best dyadic PMF for this $R$, and then update the value of $R$. The best PMF for a given $R$ is in the original formulation of the \textsc{Lec} algorithm found as follows. A subset of $\ell$ nonzero entries of $\vecp$ is chosen. A Huffman-like procedure of complexity $\mathcal{O}(m\log m)$ then finds the best dyadic PMF with $\ell$ nonzero entries. There are $m-1$ values for $\ell$ that have to be evaluated, the complexity of the overall procedure is thus roughly $\mathcal{O}(m^2\log m)$.

From \eqref{eq:dncPenalty} and a careful study of the original formulation in \cite[Sec.~III,V]{Lempel1973}, it can be shown that the iteration step is equivalent to minimizing the weighted KL-distance $\kl(\vecp\Vert\vecp^{*R})$. This can be done with complexity $\mathcal{O}(m\log m)$ by \textsc{Ghc}. A formulation of the complete \textsc{Lec} algorithm with \textsc{Ghc} as the iteration step is provided in Algorithm~\ref{alg:lec}. Besides improving the complexity of the iteration step from $\mathcal{O}(m^2\log m)$ to $\mathcal{O}(m\log m)$, our formulation answers a question that was raised in \cite{Abrahams1999}, namely how the \textsc{Lec} algorithm could be used to find the dyadic PMF that minimizes the KL-distance $\kl(\vecp\Vert\vecp^*)$. The simple answer is to perform the iteration step once with $R=1$. An implementation in MATLAB of our formulation of the \textsc{Lec} algorithm can be found at \cite{website:ghc}.

%% file: ghc.tex
\section{Optimality of \textsc{Ghc}}
Denote by $\vecx$ some non-negative vector with $m$ entries. Assume $\vecx$ is ordered, i.e., $x_1\geq x_2\geq\dotsb\geq x_m$. We now show that \textsc{Ghc} minimizes $\kl(\vecp \Vert \vecx)$ over all dyadic PMFs $\vecp$. The PMF $\vecp$ is dyadic if and only if there exist numbers $\ell_i\in\mathbb{N}$, $i=1,\dotsc,m$, such that $p_i=2^{-\ell_i},\forall i$ and $\sum_i 2^{-\ell_i}=1$. This is equivalent to $\ell_1,\dotsc,\ell_m$ being the codeword lengths of a full prefix-free code \cite[Sec. 2.3.2]{Gallager2008}. Using this, we can write
\begin{align}
\kl(\vecp \Vert \vecx)=\sum_i p_i\log\frac{p_i}{x_i}&=\log(2)\sum_i p_i\log_2\frac{p_i}{x_i}\\
&=\log(2)\sum_i 2^{-\ell_i}(-\log_2 x_i-\ell_i).
\end{align}
We define $\vecu$ by $u_i=-\log_2 x_i,\forall i$. Omitting the constant factor $\log 2$, our aim is thus to minimize
\begin{align}
\sum_i 2^{-\ell_i}(u_i-\ell_i)\label{eq:algorithmicProblem}
\end{align}
subject to $\ell_1,\dotsc,\ell_m$ are the codeword lengths of a full prefix-free code. Based on \eqref{eq:algorithmicProblem}, we now prove the optimality of \textsc{Ghc} in a way similar to the proof given in \cite[Sec. 2.5.3]{Gallager2008} for the optimality of Huffman coding.

Assume for now that an optimal algorithm assigns finite values to the codeword lengths $\ell_m$ and $\ell_{m-1}$ of the two least likely symbols, which correspond to the greatest entries $u_m$ and $u_{m-1}$ of $\vecu$. We now show that in this case, there is an optimal algorithm for which $\ell_m=\ell_{m-1}$.
\begin{lemma}\label{lem:ordered}
For an optimal algorithm, $u_i>u_j$ implies $\ell_i\geq \ell_j$. 
\end{lemma}
\begin{IEEEproof}
Assume the contrary, i.e., $u_i>u_j$ and $\ell_i < \ell_j$. Consider the term
\begin{align}
2^{-\ell_i}(u_i-\ell_i)+2^{-\ell_j}(u_j-\ell_j).
\end{align}
By interchanging $\ell_i$ and $\ell_j$, the term decreases:
\begin{align}
&[2^{-\ell_j}(u_i-\ell_j)+2^{-\ell_i}(u_j-\ell_i)]-[2^{-\ell_i}(u_i-\ell_i)+2^{-\ell_j}(u_j-\ell_j)]
\\
&\qquad\qquad\qquad
=2^{-\ell_j}(u_i-u_j)+2^{-\ell_i}(u_j-u_i)\\
&\qquad\qquad\qquad=(\underbrace{2^{-\ell_i}-2^{-\ell_j}}_{>0})(\underbrace{u_j-u_i}_{<0})<0
\end{align}
so any code with $u_i>u_j$ and $\ell_i < \ell_j$ is not optimal.
\end{IEEEproof}
\begin{lemma}\label{lem:longest}
There is an optimal algorithm for which the codewords of the two greatest entries $u_m$ and $u_{m-1}$ are siblings, i.e., $\ell_m=\ell_{m-1}$, and in addition, no other codeword is longer than $\ell_m$ and $\ell_{m-1}$. 
\end{lemma}
\begin{IEEEproof}
In a full prefix-free code, the sibling of the longest codeword is also a longest codeword. According to Lemma~\ref{lem:ordered}, if $u_m>u_{m-1}>u_{m-2}\geq\dotsb$, an optimal algorithm assigns the two longest codewords to $u_m$ and $u_{m-1}$. If only $u_m\geq u_{m-1}\geq u_{m-2}\geq\dotsb$, assigning the two longest codewords to $u_m$ and $u_{m-1}$ does  not change optimality.
\end{IEEEproof}
We can now use $\ell_m=\ell_{m-1}$ to rewrite \eqref{eq:algorithmicProblem}:
\begin{align}
\sum\limits_{i=1}^m 2^{-\ell_i}(u_i-\ell_i)&=\sum\limits_{i=1}^{m-2} 2^{-\ell_i}(u_i-\ell_i)
+2^{-\ell_{m-1}}(u_{m-1}-\ell_{m-1})+2^{-\ell_m}(u_m-\ell_m)\\
&=\sum\limits_{i=1}^{m-2} 2^{-\ell_i}(u_i-\ell_i)+
2^{-\ell_m}(u_{m-1}+u_m-2\ell_m)\\
&=\sum\limits_{i=1}^{m-2} 2^{-\ell_i}(u_i-\ell_i)+2^{-(\ell_m-1)}\Bigl[\Bigl(\underbrace{\frac{u_{m-1}+u_m}{2}-1}_{\triangleq u'}\Bigr)-(\underbrace{\ell_{m}-1}_{\triangleq \ell'})\Bigr]\\
&=\sum\limits_{i=1}^{m-2} 2^{-\ell_i}(u_i-\ell_i)+2^{-\ell'}(u'-\ell').
\end{align}
Thus, by combining $u_m$ and $u_{m-1}$ through
\begin{align}
u'=\frac{u_{m-1}+u_m}{2}-1\label{eq:updatingRuleU}
\end{align}
the size $m$ problem is reduced to a size $m-1$ problem.

The optimal algorithm may assign probability zero to the greatest entry $u_m$, which corresponds to $\ell_m=\infty$. We thus have
\begin{align}
\sum_{i=1}^m 2^{-\ell_i}(u_i-\ell_i)&=\sum_{i=1}^{m-1} 2^{-\ell_i}(u_i-\ell_i)+2^{-\infty}(u_i-\infty)
=\sum_{i=1}^{m-1} 2^{-\ell_i}(u_i-\ell_i)
\end{align}
where we used the convention $-0\log 0 = 0$ and equivalently $2^{-\infty}\infty = 0$. Thus, if $\ell_m=\infty$, the size $m$ problem reduces to a size $m-1$ problem.

It remains to check if it is better to assign probability zero to $u_m$ or to combine $u_m$ and $u_{m-1}$. First, assume the algorithm combines $u_m$ and $u_{m-1}$. Then the contribution to the sum \eqref{eq:algorithmicProblem} is $2^{-\ell'}(u'-\ell')$. We can now assign probability zero to $u_m$ and use the codeword of $u'$ for $u_{m-1}$. The contribution of $u_m$ to \eqref{eq:algorithmicProblem} is then zero and the contribution of $u_{m-1}$ is $2^{-\ell'}(u_{m-1}-\ell')$. Thus, since our aim is to minimize \eqref{eq:algorithmicProblem}, doing the former is better if and only if
\begin{align}
2^{-\ell'}(u_{m-1}-\ell')&>2^{-\ell'}(u'-\ell')\\
\Leftrightarrow (u_{m-1}-\ell')&>\Bigl(\frac{u_{m-1}+u_m}{2}-1\Bigr)-\ell'\\
\Leftrightarrow u_{m-1}&>\frac{u_{m-1}+u_m}{2}-1\\
\Leftrightarrow u_{m-1}&>u_m-2.\label{eq:choiceU}
\end{align}

Recalling $x_i=2^{-u_i}$, the updating rule \eqref{eq:updatingRuleU} and the condition \eqref{eq:choiceU} can be expressed in terms of $\vecx$ as
\begin{align}
x'=
\left\{
\begin{array}{ll}
x_{m-1},&\text{if } x_{m-1}\geq 4 x_m\\
2\sqrt{x_{m-1}x_m},&\text{if }x_{m-1}< 4 x_m.
\end{array}
\right.
\end{align}

%% file: geometric.bbl
\begin{thebibliography}{10}
\providecommand{\url}[1]{#1}
\csname url@samestyle\endcsname
\providecommand{\newblock}{\relax}
\providecommand{\bibinfo}[2]{#2}
\providecommand{\BIBentrySTDinterwordspacing}{\spaceskip=0pt\relax}
\providecommand{\BIBentryALTinterwordstretchfactor}{4}
\providecommand{\BIBentryALTinterwordspacing}{\spaceskip=\fontdimen2\font plus
\BIBentryALTinterwordstretchfactor\fontdimen3\font minus
  \fontdimen4\font\relax}
\providecommand{\BIBforeignlanguage}[2]{{%
\expandafter\ifx\csname l@#1\endcsname\relax
\typeout{** WARNING: IEEEtran.bst: No hyphenation pattern has been}%
\typeout{** loaded for the language `#1'. Using the pattern for}%
\typeout{** the default language instead.}%
\else
\language=\csname l@#1\endcsname
\fi
#2}}
\providecommand{\BIBdecl}{\relax}
\BIBdecl

\bibitem{Kschischang1993}
F.~R. Kschischang and S.~Pasupathy, ``Optimal nonuniform signaling for
  {G}aussian channels,'' \emph{{IEEE} Trans. Inf. Theory}, vol.~39, no.~3, pp.
  913--929, 1993.

\bibitem{Lempel1973}
A.~Lempel, S.~Even, and M.~Cohn, ``An algorithm for optimal prefix parsing of a
  noiseless and memoryless channel,'' \emph{{IEEE} Trans. Inf. Theory},
  vol.~19, no.~2, pp. 208--214, 1973.

\bibitem{Kerpez1991}
K.~J. Kerpez, ``Runlength codes from source codes,'' \emph{{IEEE} Trans. Inf.
  Theory}, vol.~37, no.~3, pp. 682--687, 1991.

\bibitem{Ungerboeck2002}
G.~Ungerboeck, ``Huffman shaping,'' in \emph{Codes, Graphs, and Systems},
  R.~Blahut and R.~Koetter, Eds.\hskip 1em plus 0.5em minus 0.4em\relax
  Springer, 2002, ch.~17, pp. 299--313.

\bibitem{Abrahams1998}
J.~Abrahams, ``Variable-length unequal cost parsing and coding for shaping,''
  \emph{{IEEE} Trans. Inf. Theory}, vol.~44, no.~4, pp. 1648--1650, 1998.

\bibitem{Bocherer2010e}
G.~B\"ocherer, F.~Altenbach, and R.~Mathar, ``Capacity achieving modulation for
  fixed constellations with average power constraint,'' 2010, submitted to ICC
  2011.

\bibitem{website:ghc}
G.~B\"ocherer, ``Geometric huffman coding,''
  \url{http://www.georg-boecherer.de/ghc}, Dec. 2010.

\bibitem{Cover2006}
T.~M. Cover and J.~A. Thomas, \emph{Elements of Information Theory},
  2nd~ed.\hskip 1em plus 0.5em minus 0.4em\relax John Wiley \& Sons, Inc.,
  2006.

\bibitem{Cormen2001}
T.~H. Cormen, C.~E. Leiserson, R.~L. Rivest, and C.~Stein, \emph{Introduction
  to Algorithms}, 2nd~ed.\hskip 1em plus 0.5em minus 0.4em\relax The MIT Press,
  2001.

\bibitem{Gallager2008}
R.~G. Gallager, \emph{Principles of Digital Communication}.\hskip 1em plus
  0.5em minus 0.4em\relax Cambridge University Press, 2008.

\bibitem{Gallager1968}
------, \emph{Information Theory and Reliable Communication}.\hskip 1em plus
  0.5em minus 0.4em\relax John Wiley \& Sons, Inc., 1968.

\bibitem{Krause1962}
R.~M. Krause, ``Channels which transmit letters of unequal duration,''
  \emph{Inf. Contr.}, vol.~5, pp. 3--24, 1962.

\bibitem{Marcus1957}
R.~S. Marcus, ``Discrete noiseless coding,'' Master's thesis, MIT, 1957.

\bibitem{Abrahams1999}
J.~Abrahams, ``Correspondence between variable length parsing and coding,'' in
  \emph{The mathematics of information coding, extraction and distribution},
  G.~Cybenko, D.~P. O'Leary, and J.~Rissanen, Eds.\hskip 1em plus 0.5em minus
  0.4em\relax Springer, 1999, ch.~1, pp. 1--7.

\end{thebibliography}
